\documentclass[aps,pra,nobalancelastpage,superscriptaddress,twocolumn,10pt]{revtex4-2}
\usepackage{color,ifthen,amsthm,amsmath,amsxtra,amsfonts,dsfont,graphicx,bm,tikz,scalerel,wasysym,bbm,graphicx,amsthm, braket,physics,empheq, comment}
\usepackage[colorlinks=true,linkcolor=blue, citecolor=blue, urlcolor=blue, bookmarks]{hyperref}
\usepackage[inline]{enumitem}

\usetikzlibrary{decorations.pathreplacing,angles,quotes}
\newcommand{\be}{\begin{equation}}
\newcommand{\ee}{\end{equation}}
\newcommand{\bea}{\begin{eqnarray}}
\newcommand{\eea}{\end{eqnarray}}
\newcommand{\bse}{\begin{subequations}}
\newcommand{\ese}{\end{subequations}}

\theoremstyle{plain}

\theoremstyle{plain}

\theoremstyle{plain}

\usepackage{amsthm}

\definecolor{darkGreen}{RGB}{0,110,0}
\definecolor{darkBlue}{RGB}{0,0,130}

\begin{document}
\title{Symmetry breaking in chaotic many-body quantum systems at finite temperature}
\date{\today}

 \author{Angelo Russotto}
 \affiliation{SISSA, via Bonomea 265, 34136 Trieste, Italy}
 \affiliation{INFN Sezione di Trieste, via Bonomea 265, 34136 Trieste, Italy}

 \author{Filiberto Ares}
 \affiliation{SISSA, via Bonomea 265, 34136 Trieste, Italy}
 \affiliation{INFN Sezione di Trieste, via Bonomea 265, 34136 Trieste, Italy}

 \author{Pasquale Calabrese}
 \affiliation{SISSA, via Bonomea 265, 34136 Trieste, Italy}
 \affiliation{INFN Sezione di Trieste, via Bonomea 265, 34136 Trieste, Italy}
 \affiliation{The Abdus Salam International Center for Theoretical Physics, Strada Costiera 11, 34151 Trieste, Italy}

\begin{abstract}
Recent work has shown that the entanglement of finite-temperature eigenstates in chaotic quantum many-body local Hamiltonians can be accurately described by an ensemble of random states with an internal $U(1)$ symmetry. We build upon this result to investigate the universal symmetry-breaking properties of such eigenstates. As a probe of symmetry breaking, we employ the entanglement asymmetry, a quantum information observable that quantifies the extent to which symmetry is broken in a subsystem. This measure enables us to explore the finer structure of finite-temperature eigenstates in terms of the $U(1)$-symmetric random state ensemble; in particular, the relation between the Hamiltonian and the effective conserved charge in the ensemble. Our analysis is supported by analytical calculations for the symmetric random states, as well as exact numerical results for the Mixed-Field Ising spin-$1/2$ chain, a paradigmatic model of quantum chaoticity.
\end{abstract} 

\maketitle
\textit{Introduction.---}
Recently, there has been a growing interest in characterizing typical eigenstates
of generic many-body quantum Hamiltonians. This interest is
motivated by various long-standing puzzles in 
statistical quantum mechanics, such as the 
emergence of chaos and the thermalization of 
isolated many-body quantum systems~\cite{d-91, s-94,s-99, rdo-08,dls-13,akpr-16}. At the heart 
of these questions lies random matrix theory, as in the Eigenstate Thermalization Hypothesis (ETH)~\cite{s-99,d-18}. 
The universal statistical properties of the eigenstates 
of chaotic many-body Hamiltonians are usually 
well described by ensembles of random matrices. 
For example, a defining feature of quantum chaos 
is the emergence of the Wigner-Dyson statistics 
for the energy level spacings~\cite{bgs-84, sr-10,sr2-10}. The entanglement entropy~\cite{vr-17,lg-19,ms-19,vhbr-17,ksvr-23,bhkrv-22,ypzrh-23, phfr-23, bd-19, y-19, y-21, mcl-20, hmk-22} and the spectral form factor~\cite{cdc-18,cdc2-18,klp-18,bkt-18} also follow random matrix predictions.  

However, the application of random matrix theory is generally limited to mid-spectrum eigenstates --- those with maximal entropy that correspond to infinite temperature in the thermodynamic limit. The identification of random matrix ensembles that account for finite-temperature eigenstates is an open active problem \cite{fk-19,rdsg-20,bpmgs-21}. In the novel paper~\cite{ljkr-25} (see also~\cite{rjk-24, lr-24,glhr-24}), it is shown that the entanglement entropy statistics of the finite-temperature eigenstates of local chaotic spin chains can be accurately described by pure random states that are endowed with a $U(1)$ local conserved charge. The crucial point is that energy conservation combined with the locality of interactions
induces an approximate conserved charge in the eigenstates. In this case, the charge density of the random state ensemble is related to the energy density in the spin chain.

In this work, using the setup of Ref.~\cite{rjk-24}, we show that the same $U(1)$-symmetric random state ensemble further describes the symmetry-breaking properties of finite-temperature eigenstates. To this end, we employ the entanglement asymmetry, a new observable based on entanglement entropy that measures how much a symmetry is broken in a subsystem. The entanglement asymmetry is a useful tool to 
monitor the time evolution of (broken) symmetries after quantum quenches and observing the quantum 
Mpemba effect~\cite{amc-23, rylands-24, sakc-24, k-24, rvc-24, carc-24, yac-24, fac-24, yca-24, avm-24, cma-24,cv-24, khor-24, bds-24, liu-24-2, teza-25, acm-25}, including random circuits~\cite{tcd-24,krb-24,lzyz-24,fcb-25,ylz-25, amcp-25} and experiments~\cite{joshi-24}. It has also been studied in field theories~\cite{chmp-20, chmp-21, cc-23, bgs-24, fadc-24, frc-24, kmop-24, cm-23, lmac-25}, Haar random states~\cite{ampc-24, rac-24, ct-25}, and, from a different perspective, in quantum information resource theory~\cite{vawj-08, gms-09,tfhtp-24}. 

\begin{figure}[t]
\includegraphics[width=0.49\textwidth]{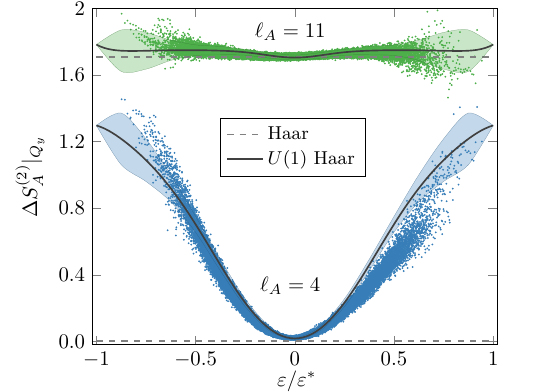}
\centering
\caption{R\'enyi$-2$ asymmetry for the charge $Q_y$ of the eigenstates of the mixed-field Ising Hamiltonian~\eqref{model} with $L = 16$ sites in the chaotic point $g = 1.1, h = 0.35$ for a subsystem of length $\ell_A = 4$ (blue points) and $\ell_A = 11$ (green points). The energy density $\varepsilon$ of the eigenstates is rescaled by $\varepsilon^* = 1.54$. The black continuous lines are the analytical prediction~\eqref{eq:EDS} obtained using the ensemble of $U(1)$-symmetric Haar random states, with conserved charge orthogonal to $Q_y$. The blue and green shaded regions correspond to the confidence interval $\mathbb{E}[\Delta S^{(2)}_A]\pm 3 \sigma$, where the variance $\sigma^2$ is estimated numerically by sampling the ensemble of $U(1)$-symmetric random states. The grey dashed lines are the prediction~\eqref{eq:st_haar} of the standard Haar unitary ensemble, which is expected to be valid only for $\varepsilon\sim0$.}
\label{fig:asymfullspectrum} 
\end{figure}

The entanglement asymmetry neatly detects the effective conserved $U(1)$ charge induced in the energy eigenstates by the locality of interactions. This allows us to determine the charge that generates the $U(1)$ symmetry of the ensemble.  Moreover, we study the symmetry-breaking for other $U(1)$ local charges and compare with the prediction of the symmetric random states. Fig.~\ref{fig:asymfullspectrum} summarizes our main results. The symbols are the entanglement asymmetry of all the eigenstates of the chaotic Mixed-Field Ising spin-$1/2$ chain for a non-conserved charge orthogonal (in the sense later specified) to the one that is effectively conserved. Their asymmetry is calculated numerically in a chain of $L=16$ spins for two different subsystems of $\ell_A=4$ and $11$ contiguous spins. The average of this asymmetry in the $U(1)$-symmetric random state ensemble can be calculated analytically and its prediction is represented by the continuous curves in Fig.~\ref{fig:asymfullspectrum}.
We find a remarkable agreement. In contrast, the prediction (dashed lines) using the standard Haar random state ensemble, which does not have any symmetry, only captures the behavior of mid-spectrum eigenstates.

In the rest of the paper, we first briefly review the main ideas and setup of Ref.~\cite{rjk-24}, then introduce the entanglement asymmetry, and finally study it in both the $U(1)$-symmetric random state ensemble and the Hamiltonian system, comparing the two.  

\textit{Eigenstates of local chaotic Hamiltonians.---} Let us consider 
a one-dimensional many-body quantum system that can be divided into two spatial regions $A$ and $B$. The system is described by a locally interacting Hamiltonian $H$, i.e. it can be partitioned as $H = H_A + H_B + H_{AB}$, where $H_A\,( H_B)$ only acts on $A\,(B)$ and $H_{AB}$ contains all of the terms coupling $A$ and $B$. Any eigenstate $\ket{\psi_E}$ of $H$ with energy $E$ then admits the decomposition $\ket{\psi_E} = \sum_{ij} c_{ij} \ket{\psi_{E_i}}_A\otimes\ket{\psi_{E_j}}_B$, where $\ket{\psi_{E_i}}_{A(B)}$ is an eigenstate of $H_{A(B)}$ with energy $E_{i}$. A key observation, first analyzed in \cite{d-91}, is that the locality of the interaction $H_{AB}$ makes the matrix $c_{ij}$ banded; that is,  $c_{ij}$ is non-zero only for energies $E_i+E_j \in [E-\delta/2, E+\delta/2]$, where $\delta$ is set by the interaction term $H_{AB}$. The locality of the interactions implies that $\delta \sim O(1)$. More rigorously, Ref.~\cite{akl-16} proved that $c_{ij}$ are upper bounded by a decreasing exponential in $E_i + E_j - E$.
Refs.~\cite{lg-19,ms-19} used this idea to construct states consistent with the subsystem ETH~\cite{dll-18} and that describe universal properties of finite-temperature eigenstates of chaotic Hamiltonians. 
A further approximation assumed in Refs.~\cite{rjk-24,ljkr-25} is the coarse-graining of the spectrum of $H_{A(B)}$ with the scale $\delta$ by assigning to all the eigenstates in the window $\varepsilon_k-\delta/2\leq E_i\leq \varepsilon_k+\delta/2$ the same energy $\varepsilon_k$. This identification effectively discretizes the spectrum and creates degenerate energy subspaces $\mathcal{H}_{A(B)}(\varepsilon_k)$. As a result, the typical eigenstates can be approximately written as $\ket{\psi_E} \approx \sum_k c_{\varepsilon_k} \ket{\psi_{\varepsilon_k}} $ with $\ket{\psi_{\varepsilon_k}} \in \mathcal{H}_A(\varepsilon_k) \otimes \mathcal{H}_B(E-\varepsilon_k)$, akin the structure of states with a global $U(1)$ symmetry. Based on this approximation, Refs.~\cite{rjk-24,ljkr-25} conjecture that the properties of typical eigenstates of local chaotic Hamiltonians are captured by random states subject to this symmetry constraint.   

As shown in \cite{ljkr-25}, this $U(1)$-symmetric random state ensemble not only gives a better prediction than the Haar ensemble of the entanglement entropy at zero energy density but it is also able to capture the universal entanglement properties of finite-temperature eigenstates. These eigenstates have typically less entanglement entropy than the infinite-temperature one. This effect can be modeled by changing in the ensemble the sector of the total conserved charge. After introducing this $U(1)$-symmetric ensemble, we describe this correspondence in detail.

\textit{$U(1)$-symmetric random state ensemble.---}
The ensemble of $U(1)$-symmetric random states employed in~\cite{rjk-24,ljkr-25} was initially studied in~\cite{y-19,y-21}. A complete characterization of the first two moments of the entanglement entropy has been given in~\cite{bd-19} and of its symmetry-resolution in~\cite{mcp-22, lau-22, ghasemi-25}. To define it, we need to consider a charge operator $Q$ that generates a global $U(1)$ group.  The Hilbert space of the full system can be then decomposed as the direct sum of each charge sector, $\mathcal{H} = \oplus_M \mathcal{H}(M)$, where $M$ is an eigenvalue of $Q$. If the total charge is the sum of the contributions of $A$ and $B$, $Q=Q_A+Q_B$, then the subspace $\mathcal{H}(M)$ of fixed total charge $M$ decomposes as 
\begin{equation}\label{eq:decomposition_HM}
     \mathcal{H}(M) = \bigoplus_{q} \mathcal{H}_A(q) \otimes \mathcal{H}_B(M-q).
\end{equation}
Thus, the $U(1)$-symmetric states $\ket{\Psi(M)} \in \mathcal{H}(M)$, i.e. those satisfying $Q\ket{\Psi(M)}=M\ket{\Psi(M)}$, can be written as $\ket{\Psi(M)} = \sum_q \sqrt{p_q} \ket{\psi_q}$, where $\ket{\psi_q} \in \mathcal{H}_A(q) \otimes \mathcal{H}_B(M-q)$. The corresponding reduced density matrix that describes the subsystem $A$ is of the form $\rho_A = \sum_q p_q \rho_A(q)$. Here $p_q$ is the probability of finding a charge $q$ in $A$.
In Ref.~\cite{bd-19}, it is proved that the uniform measure over $\mathcal{H}(M)$ subject to the symmetry constraint~\eqref{eq:decomposition_HM} is equal to the product of the uniform measure in each subspace $\mathcal{H}_A(q) \otimes \mathcal{H}_B(M-q)$, which is the usual Haar unitary measure, times the probability measure over the coefficients $\{p_q\}$, which is given by the Dirichlet distribution with parameters $d_q = d_{A,q} d_{B,q}$, where $d_q$ is the dimension of the space $\mathcal{H}_A(q) \otimes \mathcal{H}_B(M-q)$ for each possible $q$.

\begin{figure*}[t]
\centering
\includegraphics[width=0.99\textwidth]{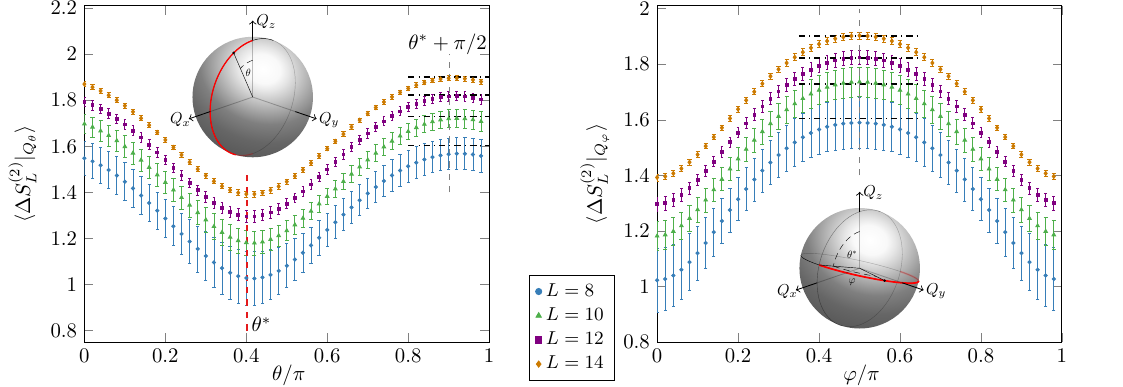}
  \caption{\label{fig:asymvsthetaphi} 
  Left panel. The symbols are the average full system Rényi-2 asymmetry for the charge $Q_{\theta}$ in a window of mid-spectrum eigenvectors of the Hamiltonian~\eqref{model} centered at $\varepsilon=0$ (we take $46,68,100,200$ states for $L = 8,10,12,14$, respectively) as a function of the angle $\theta$. This family of charges are represented by the red arc in the inset sphere. The errorbars are the standard deviation of the asymmetry over the chosen sample of eigenstates. Independently on $L$, the asymmetry shows a minimum at a value of $\theta$ very close to $\theta^*=\arctan(g/h)$, as highlighted by the red dashed line. 
  The black dash-dotted lines are the analytical prediction~\eqref{eq:EDS} for $\mathbb{E}[\Delta S^{(2)}_L]$ for a charge orthogonal to the one conserved in the $U(1)$-symmetric random state ensemble in the sector $M=L/2$. Right panel. Same analysis but for charges belonging to the arc connecting $Q_{\theta^*}$ and $Q_y$ (red curve in the sphere inset). The maximum is reached at $\varphi \approx \pi/2$ and is again compatible with the prediction~\eqref{eq:EDS} with $M=L/2$.}
\end{figure*}

\textit{Model.---}
As a concrete example, we consider the one-dimensional Mixed-Field Ising Model (MFIM), a paradigmatic model of quantum chaos~\cite{zkh-15,sr-10,dls-13}. This is a spin-$1/2$ chain with Hamiltonian:
\begin{equation}
\label{model}
    H = \sum_{j} (\sigma^z_j\sigma^z_{j+1}+ g \sigma_j^x + h \sigma_j^z),
\end{equation}
where $\sigma^{x,y,z}_j$ are the Pauli matrices on the $j$-th site. We take a chain with $L$ sites and open boundary conditions. Following Ref.~\cite{rjk-24,ljkr-25}, we include two boundary fields $h_1 \sigma^z_1$, $h_L \sigma^z_L$ to break the inversion symmetry under the spatial reflection of the chain. In this way, the
Hamiltonian does not possess any symmetry, and when the couplings are both $O(1)$, the model is far enough away from any possible integrability regime. More specifically, we set $g = 1.1$ and $h = 0.35$. These values belong to a commonly used range of parameters to study quantum chaos \cite{bch-11,kh-13,zkh-15} and they correspond to the most chaotic point of the model~\eqref{model} in the sense discussed in \cite{rjk-24}.

We now make explicit the connection between the energy of the eigenstates and the charge of $U(1)$-symmetric random states. The only requirement of the charge $Q$ that generates the symmetry is to satisfy $Q = Q_A + Q_B$. Under this assumption, we can consider, without loss of generality, a charge that takes values $0$ or $1$ at each spin. In that case, the charge density $s\equiv M/L$, defined in the interval $0\leq s\leq 1$, can be identified 
with the energy density $\varepsilon\equiv E/L$ of the eigenstates of $H$ by~\cite{ljkr-25}
\begin{equation}\label{eq:correspondence}
\varepsilon/\varepsilon^* = 2 s -1, 
\end{equation}
such that the charge sector with largest degeneracy, $s=1/2$, corresponds to the zero energy density, $\varepsilon=0$. 
The rescaling factor $\varepsilon^*$ is determined by the scale of the energy fluctuations, $\varepsilon^*=\sqrt{\langle H^2\rangle/L}$. 
%Let $M \equiv s L$ be the total conserved charge, we thus have $0\leq s \leq 1$. We consider the rescaled energy density $\varepsilon_n/\varepsilon^* = E_n/(L\, \varepsilon^*)$, where $\varepsilon^*$ is given by the scale of energy fluctuations set by the energy variance as $\varepsilon^* = \sqrt{\langle H^2\rangle/L}$.
%The relation between the charge sector and the rescaled energy density is $\varepsilon/\varepsilon^* = 2 s -1$, such that for zero energy density we have the sector with maximal dimension. 
At finite-size, Ref.~\cite{ab-14} showed that the density of states peaks around an energy $E_p \neq 0$ that tends to zero when $L\to\infty$. For this reason, we estimate $E_p$ and $\varepsilon^*$ by fitting the density of states obtained by exactly diagonalizing~\eqref{model}. We then consider as rescaled energy density $\varepsilon/\varepsilon^* = (E-E_p)/(L\, \varepsilon^*)$.

However, we have not provided a specific expression for the conserved charge in the ensemble of $U(1)$-symmetric random states. In our particular example, it should correspond to the approximate global $U(1)$-symmetry induced by energy conservation and the locality of the interactions in the eigenstates of the Hamiltonian~\eqref{model}. To further understand and identify the explicit form of the charge that generates this effective symmetry, a convenient tool is the entanglement asymmetry.

\textit{Entanglement asymmetry.---}
The entanglement asymmetry quantifies the extent to which a symmetry transformation generated by a charge $Q_A$ is broken within a subsystem $A$. Its definition requires the introduction of the symmetrized reduced density matrix $\rho_{A,Q} = \sum_q \Pi_q \rho_A \Pi_q$, where $\Pi_q$ is projector on the eigenspace of $Q_A$ corresponding to the eigenvalue $q$. The Rényi-$n$ entanglement asymmetry is then defined as the difference between the Rényi-$n$ entropies $S_n(\rho) = \frac{1}{1-n}\log \text{Tr}[\rho^n]$ of $\rho_{A,Q}$ and $\rho_A$,
\begin{equation}\label{eq:asymm}
    \Delta S^{(n)}_A = S_n(\rho_{A,Q})-S_n(\rho_A).
\end{equation}
This quantity is non-negative $\Delta S^{(n)}_A \geq 0$ and $\Delta S^{(n)}_A = 0$ if and only if $[\rho_A,Q_A] = 0$. Here, we restrict us to $n=2$, although we expect the same qualitative behavior for any $n$.

\textit{Analytical results.---} \label{an_res}
Since the effective conserved charge must be local, we restrict ourselves to the family of $U(1)$-charges $Q_{\hat{n}}=\sum_{\alpha} \hat{n}_\alpha Q_\alpha$, where $Q_\alpha=\sum_j \sigma_j^\alpha$, with $\alpha=x, y, z$ and $\hat{n}\in \mathbb{R}^3$ satisfying $|\hat{n}|=1$. Let us assume that, for a specific $\hat{n}^*\in\mathbb{R}^3$, $Q_{\hat{n}^*}$ is the charge that generates the $U(1)$-symmetry in the ensemble of random states. In that case, the asymmetry of the random states is $\Delta S_A^{(2)}|_{\hat{n}^*}=0$, while, for any other $\hat{n}$, we will typically have $\Delta S_A^{(2)}|_{\hat{n}\neq \hat{n}^*}\neq 0$. In particular, we consider a charge $Q_{\perp}\equiv Q_{\hat{n}_\perp}$ orthogonal to $Q_{\hat{n}^*}$, in the sense that $\hat{n}_\perp\cdot \hat{n}^*=0$. For this charge, the average asymmetry $\mathbb{E}[\Delta S_A^{(2)}|_{Q_\perp}]$ over the ensemble of $U(1)$-symmetric random states in the sector $M$ has the following analytic expression. The ratio $R\equiv \mathbb{E}[\text{Tr}[\rho_{A,Q_{\perp}}^2]]/\mathbb{E}[\text{Tr}[\rho_{A}^2]]$ can be exactly computed using Weingarten calculus~\cite{w-78, cs-06, pv-22, fknv-23} (see derivation in~\cite{sm}) and reads
\begin{equation}
\label{eq:R}
    R = \frac{2^{-2 \ell_A} \binom{2 \ell_A}{\ell_A} \sum_j d_{B,j} d_{A,j}^2 + \chi(L,\ell_A,M)}{\sum_j d_{B,j} d_{A,j}^2+\sum_j d_{B,j}^2 d_{A,j}},
\end{equation}
where $d_{A,j} = \binom{\ell_A}{j}$, $d_{B,j} = \binom{L-\ell_A}{M-j}$,
\begin{multline}
\label{eqR_chi}
    \chi(L,\ell_A,M)  =\\=2^{-\ell_A} \sum_{m = 0}^{\ell_A} \frac{2^{-2m}(2 m)!}{(m!)^2}\binom{L-2m}{M}^2\binom{\ell_A}{2 m} \mathcal{F}(m,L,M)^2,
\end{multline}
and $\mathcal{F} (m,L,M)$ is the hypergeometric function ${}_2F_1$ with parameters $\mathcal{F}(m,L,M) \equiv {}_2F_1(-2m,-M,1-2m+L-M;-1)$. 

We now further assume the self-averaging approximation $ \mathbb{E}[\log \text{Tr}[\rho_{A,Q}^2]]\simeq \log \mathbb{E}[ \text{Tr}[\rho_{A,Q}^2]]$. This property can be proven for the Haar ensemble \cite{kf-21,rac-24} when $L\to\infty$. For the $U(1)$ ensemble under consideration, we lack of an analytical proof, although the numerical analysis performed in~\cite{sm} supports it, even at finite $L$. Applying the self-averaging property, we obtain
\begin{equation}
\label{eq:EDS}
    \mathbb{E}[\Delta S^{(2)}_A|_{Q_{\perp}}] \simeq - \log R. 
\end{equation}
A numerical check of this result is presented in~\cite{sm}. Eq.~\eqref{eq:EDS} has a non-trivial dependence on $\ell_A$ and the total conserved charge $M$ of the random states. This makes it drastically different from the result for the ordinary Haar ensemble of states, calculated in Refs.~\cite{ampc-24,rac-24}. In that case, any symmetry is broken and, for any charge $Q_{\hat{n}}$, the average asymmetry is
\begin{equation}\label{eq:st_haar}
\mathbb{E}[\Delta S_A^{(2)}|_{\hat{n}}]=-\log\left[\frac{1}{2^{2\ell_A-L}+1}\left(1+2^{-L}\frac{(2\ell_A)!}{(\ell_A!)^2}\right)\right].
\end{equation}

Unfortunately, the analytical analysis of the higher cumulants of $\Delta S^{(2)}_A$, and in particular the variance, is much more complicated and we are not able to obtain an analytical prediction. For this reason, the variance displayed in Fig.~\ref{fig:asymfullspectrum} is obtained numerically.

In what follows, we will show that Eq.~\eqref{eq:EDS} captures the symmetry-breaking properties of the eigenstates of the chaotic Hamiltonian~\eqref{model}.

\textit{Numerical results.---} 
\label{num_res}
Let us first focus on the Rényi-$2$ asymmetry of the full system ($\ell_A=L$) at zero mean energy density ($\varepsilon=0$). We consider a statistically significant number of mid-spectrum eigenstates centered around zero energy and compute numerically its average asymmetry. We denote the average over eigenstates by $\langle \cdot \rangle$, while we keep the notation $\mathbb{E}[\cdot]$ for the expectation value over the random states.
In the left panel of Fig.~\ref{fig:asymvsthetaphi}, we show the results, taking $Q_\theta=\cos\theta \,Q_z+\sin\theta\, Q_x$  as the charge in Eq.~\eqref{eq:asymm}, for various systems sizes $L$. We observe that $\langle \Delta S_A^{(2)}|_{Q_\theta}\rangle$ is smooth in $\theta$ and shows a minimum close to $\theta^*=\arctan(g/h)$, while it reaches a maximum for the orthogonal charge $Q_{\theta^*+\pi/2}$. The dashed horizontal lines represent the values predicted by Eq.~\eqref{eq:EDS}  for the asymmetry of a charge orthogonal to the conserved one in the random ensemble
in the sector $M=L/2$ (which corresponds to $\varepsilon = 0$ Eq.~\eqref{eq:correspondence}).   

The same analysis can be repeated for a charge $Q_\varphi$ obtained rotating $Q_{\theta^*}$ by an angle $\varphi$ around the direction orthogonal to it in the $x-z$ plane (see the inset of Fig.~\ref{fig:asymvsthetaphi} right). There is a minimum at $\varphi=0$, which corresponds to $Q_{\theta^*}$, and a maximum at the orthogonal direction. The maximum is again well-captured by Eq.~\eqref{eq:EDS} taking $M=L/2$ and $\ell_A=L$. From both plots in Fig.~\ref{fig:asymvsthetaphi}, we conclude that the approximate conserved charge in the model~\eqref{model} corresponding to the $U(1)$-symmetry of the random states is $Q_{\theta^*}=gQ_x+hQ_z$; these are the on-site terms of the Hamiltonian~\eqref{model}. Of course, the value of the minimum is not comparable to zero for any value of $L$ analyzed. This is not surprising, as we know that $H$ does not have exact symmetries. However, the existence of a well-defined single minimum supports the description of the infinite-temperature eigenstates of~\eqref{model} in terms of $U(1)$-symmetric random states in the sector $M=L/2$. Moreover, the asymmetry of these states with respect to an orthogonal charge is well captured by the $U(1)$-ensemble. According to the correspondence~\eqref{eq:correspondence}, Eq.~\eqref{eq:EDS} should also describe finite-temperature eigenstates ($\varepsilon\neq 0$) when taking $M<L/2$, as shown in Fig.~\ref{fig:asymfullspectrum}. We observe that, although the agreement is generally good, the numerical data are slightly not even in $\varepsilon$ contrary to our theoretical prediction.
It would be interesting to characterize and understand this small subleading correction. 

\begin{figure}[t]
\includegraphics[width=0.49\textwidth]{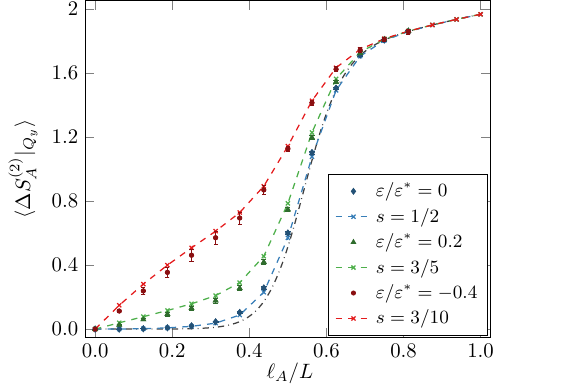}
\centering
\caption{Average R\'enyi-2 asymmetry over a window of eigenstates of the Hamiltonian \eqref{model} with $L=16$ centered around different rescaled energies $\varepsilon/\varepsilon^*$ as a function of $\ell_A$. We consider energy windows $\Delta \varepsilon$ of $0.002 \varepsilon^*, 0.0025 \varepsilon^*, 0.006 \varepsilon^*$ for the mean ratios $\varepsilon/\varepsilon^* = 0, 0.2 , -0.4$, respectively, to encompass in all the cases $\sim 200$ eigenstates. The error bars are the standard deviation of the asymmetry in each energy window. The crosses joined by a dashed line are the prediction of Eq.~\eqref{eq:EDS} for the $U(1)$ random state ensemble (the line is a visual guide). The relation between the energy density $\varepsilon$ and the sector $s \equiv M/L$ is reported in Eq.~\eqref{eq:correspondence}. The black dash-dotted line is the prediction~\eqref{eq:st_haar} for the standard non-symmetric Haar ensemble.}
\label{fig:DSvslA} 
\end{figure}

In Fig.~\ref{fig:DSvslA}, we further analyze the results in Fig.~\ref{fig:asymfullspectrum}. Here the symbols are the average eigenstate entanglement asymmetry for $Q_y$ over energy windows with different mean rescaled energy density $\varepsilon/\varepsilon_*$. For sufficiently small window and statistically significant number of eigenstates inside the window, this average is an estimate of the microcanonical expectation value of the entanglement asymmetry at finite temperature. The results in Fig.~\ref{fig:DSvslA} are in good agreement with the analytical prediction in Eq.~\eqref{eq:EDS} (coloured dashed lines), which clearly captures the dependence on the subsystem size. However, when moving away from the mid-spectrum, i.e. when $|\varepsilon/\varepsilon^*|\to 1$, the deviations evidence the presence of a non-universal $O(1)$ correction which is beyond the scope of our random matrix description. The black dash-dotted line is the prediction~\eqref{eq:st_haar} for a standard Haar random state, with no symmetries. As already noted in Ref.~\cite{rjk-24} for the entanglement entropy, Eq.~\eqref{eq:EDS} with $M = L/2$ provides a more accurate description of the mid-spectrum.

\textit{Conclusions.---} 
In this paper, we demonstrated the power of entanglement asymmetry in understanding the universal properties of finite-temperature eigenstates of local chaotic many-body Hamiltonians. Using the asymmetry, we determined the effective conserved charge in the energy eigenstates resulting from the locality of interactions. As we showed, this charge generates the $U(1)$ symmetry in the ensemble of random states that describes the statistical behavior of the finite-temperature eigenstates. We found that it corresponds to the on-site part of the chaotic Hamiltonian. The explicit form of this charge is essential for extracting predictions from the $U(1)$-symmetric random state ensemble. For instance, it allows us to calculate the asymmetry with respect to other non-conserved $U(1)$ charges, as illustrated in this work. The ensemble of random states endowed with this conserved charge well captures the entanglement asymmetry over the whole range of energy density of the spectrum. 

Our results only rely on the locality of the Hamiltonian interactions. Therefore, we expect that they are universal for chaotic local spin Hamiltonians. In~\cite{sm}, we present some numerical results supporting this conjecture for other Hamiltonians with the same on-site terms as~\eqref{model}, but different interactions.

Interesting avenues for future research include the analysis of the universal properties of asymmetry in chaotic models where the appropriate reference random state ensemble contains multiple commuting $U(1)$ charges, as in Ref.~\cite{lr-24}, or in systems with non-Abelian symmetries~\cite{phfr-23,mly-23,bdk-24}. It would also be compelling to explore other applications of our results for the $U(1)$-symmetric random state ensemble, such as in the characterization of the stationary state of highly chaotic dynamics~\cite{glhr-24}.

\textit{Acknowledgments.---} We thank Mario Collura, Viktor Eisler, and Colin Rylands for useful discussions. PC and FA acknowledge support from ERC under Consolidator Grant number 771536 (NEMO) and from European Union - NextGenerationEU, in the framework of the PRIN Project HIGHEST no. 2022SJCKAH$\_$002. Numerical computations were performed using the SISSA cluster Ulysses.

\onecolumngrid
\newpage 
\newcounter{equationSM}
\newcounter{figureSM}
\newcounter{tableSM}
\stepcounter{equationSM}
\setcounter{equation}{0}
\setcounter{figure}{0}
\setcounter{table}{0}
\setcounter{section}{0}
\makeatletter
\renewcommand{\theequation}{\textsc{sm}-\arabic{equation}}
\renewcommand{\thefigure}{\textsc{sm}-\arabic{figure}}
\renewcommand{\thetable}{\textsc{sm}-\arabic{table}}

\begin{center}
  {\large{\bf Supplemental Material for\\ ``Symmetry breaking in chaotic many-body quantum systems at finite temperature''}}
\end{center}
Here we report some useful information complementing the main text. In particular,
\begin{itemize}
 \item[-] In Sec.~\ref{sec:AnalyticsDeriv}, we show in detail the derivation of Eq.~(5).
  \item[-] In Sec.~\ref{sec:AddNumerics}, we provide a numerical check of the analytical results and some further numerical results.
\end{itemize}

\section{Analytical derivation of Eq.~(5)}\label{sec:AnalyticsDeriv}

Equation~(5) is the average entanglement asymmetry of the $U(1)$-symmetric Haar random ensemble with respect to a charge orthogonal to the conserved charge that generates the symmetry of this ensemble. For simplicity, we take here as conserved charge $Q_z = \sum_{j=1}^L \sigma^z_j$ and as orthogonal charge 
$Q_x = \sum_{j=1}^L \sigma^x_j$. Since by construction any state in the ensemble commutes with $Q_z$, which is the charge generating the rotations along the $z-$axis, we have that the asymmetry for $Q_x$ is the same as that for $Q_y$. Therefore, we also have that, on average, $\mathbb{E}[\Delta S^{(2)}|_{Q_x}] =\mathbb{E}[ \Delta S^{(2)}|_{Q_y}]$.
From now on, and for simplicity, we consider as conserved charge $Q'_z = \frac{1}{2}(Q_z + L)$ such that the eigenvalues are integers $q' = 0,1,2, \dots, L$. This will not modify our final result: the asymmetry is invariant under this change since the projectors in the eigenspaces of $Q_z'$ and $Q_z$ are the same. Let us take a state $\ket{\Psi(M)}$ belonging to the ensemble of $U(1)$-symmetric Haar random states with fixed total charge $Q_z'\ket{\Psi(M)}=M\ket{\Psi(M)}$. We divide the full system in a subsystem $A$ of $\ell_A$ sites and the complement $B$ with $L-\ell_A$ sites. To compute the asymmetry of the reduced density matrix $\rho_A=\Tr(\ket{\Psi(M)}\bra{\Psi(M)})$ with respect to $Q_x$, we need to calculate its symmetrization, 
\begin{equation}\label{eq:symm_Qx}
    \rho_{A,Q_x} = \sum_{k} \Pi^{(x)}_k \rho_A \Pi^{(x)}_k,
\end{equation}
where $\Pi_q^{(\alpha)}$ is the projector onto the eigenspace of with eigenvalue $q$ of $Q_{x, A}$, i.e. the restriction to $A$ of $Q_x$.
To enlighten the notation, we omit in the projectors the subscript $A$ indicating that they act on the Hilbert space $\mathcal{H}_A$. 

To exploit the symmetry properties of $\rho_A$, we find convenient to write the projectors $\Pi^{(x)}_q$ in terms of the projectors $\Pi_q^{(z)}$ of the conserved charge. Due to the algebra of Pauli matrices, they are simply related by
\begin{equation}\label{eq:proj_X_Z}
    %\Pi^{(x)}_k = e^{-i \frac{\pi}{4} Q_{y,A}} \Pi^{(z)}_k e^{i \frac{\pi}{4} Q_{y,A}}.
    \Pi^{(x)}_k = \mathcal{U}^\dagger \Pi^{(z)}_k \mathcal{U},
\end{equation}
where $\mathcal{U} \equiv \exp(i \pi Q_{y,A}/4)$. Combining Eqs.~\eqref{eq:symm_Qx} and~\eqref{eq:proj_X_Z}, the second moment of the symmetrized density matrix is then
\begin{equation}
\label{secmom}
    \text{Tr}[\rho_{A,Q_x}^2] =\sum_k \text{Tr}[\mathcal{U}^{\dagger} \Pi^{(z)}_k \mathcal{U} \rho_A \mathcal{U}^{\dagger} \Pi^{(z)}_k \mathcal{U} \rho_A].
\end{equation}
The next step is to write the operator $\mathcal{U}^{\dagger} \Pi^{(z)}_k \mathcal{U}$ in the basis of the eigenstates $\{\ket{qm}\}$ of $Q'_{z,A}$ as follows
\begin{equation}
\label{projx}
    \mathcal{U}^{\dagger} \Pi^{(z)}_k \mathcal{U} = \sum_{\substack{qq'\\ mm'}} A(k)_{qq'}^{mm'} \ket{qm}\bra{q'm'}.
\end{equation}
The state $\ket{qm}$ satisfies $Q'_{z,A} \ket{qm} = q \ket{qm}$ and $m$ labels the degenerate states in the eigenspace with eigenvalue $q$. The coefficients $A(k)_{qq'}^{mm'}$ are defined as
\begin{equation}
    A(k)_{qq'}^{mm'} \equiv \bra{qm} \Pi^{(x)}_k\ket{q'm'}.
\end{equation}
Applying Eq.~\eqref{eq:proj_X_Z} and $\Pi^{(z)}_k = \sum_p \ket{kp}\bra{kp}$,
\begin{equation}
     A(k)_{qq'}^{mm'}= 
     %\bra{qm} e^{- i \frac{\pi}{4} Q_{y,A}} \Pi^{(z)}_k  e^{ i \frac{\pi}{4} Q_{y,A}}\ket{q'm'} 
     \bra{qm} \mathcal{U}^\dagger \Pi^{(z)}_k  \mathcal{U}\ket{q'm'} 
     = \sum_{p} C_{qk}^{mp} C_{q'k}^{m'p\,*},
\end{equation}
where $C_{qq'}^{mm'}$ is the matrix element 
%$C_{qq'}^{mm'} \equiv \bra{qm} e^{- i \frac{\pi}{4} Q_{y,A}} \ket{q'm'}$
$C_{qq'}^{mm'} \equiv \bra{qm} \mathcal{U}^\dagger \ket{q'm'}$.
Plugging \eqref{projx} into \eqref{secmom}, we obtain
\begin{equation}
\label{secmom2}
        \text{Tr}[\rho_{A,Q_x}^2] = \sum_k \sum_{\substack{qq'\\ mm'}} \sum_{\substack{\Tilde{q}\Tilde{q}'\\ \Tilde{m}\Tilde{m}'}} A(k)_{qq'}^{mm'} A(k)_{\Tilde{q}\Tilde{q}'}^{\Tilde{m}\Tilde{m}'} \bra{q'm'} \rho_A \ket{\Tilde{q} \Tilde{m}} \bra{\Tilde{q}'\Tilde{m}'} \rho_A \ket{q m}.
\end{equation}
We use now the symmetry property of the density matrix, which by construction can be decomposed as $\rho_A = \sum_{q = 0}^{\ell_A} p_q \rho_{A}(q)$ where we are now summing over the eigenvalues $q$ of $Q_{z,A}'$. Using this block decomposition, the matrix elements appearing in Eq.~\eqref{secmom2} simplify as follows 
\begin{equation}
\bra{q'm'} \rho_A \ket{\Tilde{q} \Tilde{m}} = \sum_j p_j \bra{q'm'} \rho_A(j) \ket{\Tilde{q} \Tilde{m}} = \sum_j p_j \delta_{q'j} \delta_{j\Tilde{q}} \, \rho_A(j)_{m'\Tilde{m}}. 
\end{equation}
Applying this result in Eq.~\eqref{secmom2}, we finally obtain for $\text{Tr}[\rho_{A,Q_x}^2]$
\begin{equation}
        \text{Tr}[\rho_{A,Q_x}^2] = \sum_k \sum_{j,j'} \sum_{\substack{mm' \\ \Tilde{m} \Tilde{m}'}} p_j p_{j'} A(k)_{j'j}^{mm'} A(k)_{jj'}^{\Tilde{m}\Tilde{m}'} \rho_A(j)_{m' \Tilde{m}} \rho_A(j')_{\Tilde{m}' m}.
\end{equation}
This expression can be rewritten in the form
\begin{equation}
        \text{Tr}[\rho_{A,Q_x}^2] = \sum_k \sum_{j,j'} p_j p_{j'} \text{Tr}[A(k)_{j'j} \rho_A(j) A(k)_{jj'} \rho_A(j')],
\end{equation}
where the trace is implied to be taken in the subspace $\mathcal{H}_A(j')$, i.e. the eigenspace of $Q'_{z,A}$ with eigenvalue $j'$.

We are now ready to compute the expectation value over the ensemble of $U(1)$-symmetric Haar random states. Taking into account that $p_q$ and $\rho_A(q)$ are independent stochastic variables, their joint expectation value factorizes as
\begin{equation}
\label{EmomG2}
    \mathbb{E}[\text{Tr}{\rho_{A,Q_x}^2}] = \sum_k \sum_{j,j'} \mathbb{E}[ p_j p_{j'}] \, \mathbb{E}[\text{Tr}[A(k)_{j'j} \rho_A(j) A(k)_{jj'} \rho_A(j')]]. 
\end{equation}

The $p_q$ are described by the Dirichlet distribution, defined by the  measure
\begin{equation}
\label{pdf_pq}
    d \mu \left(\{p_q\}_q\right) \propto \delta\left(\sum_q p_q -1\right) \prod_q p_q^{d_q-1} dp_q.
\end{equation} 
Using it, we can compute the average $\mathbb{E}[p_j p_{j'}]$,
\begin{equation}\label{eq:av_p_q_2}
\mathbb{E}[p_j p_{j'}]=\frac{d_j d_{j'} + \delta_{jj'} d_j}{d_M(d_M+1)},
\end{equation}
 where $d_j = d_{A,j} d_{B,j}=\binom{\ell_A}{j}\binom{L-\ell_A}{M-j}$ and $d_M = \sum_{j=0}^{\ell_A} \binom{\ell_A}{j}\binom{L-\ell_A}{M-j} = \binom{L}{M}$. 
Inserting it in Eq.~\eqref{EmomG2}, we have
\begin{equation}\label{eq:av_symm_Q_X}
\mathbb{E}[\text{Tr}{\rho_{A,Q_x}^2}]= \sum_k \sum_{j,j'}\frac{d_j d_{j'} + \delta_{jj'} d_j}{d_M(d_M+1)} \, \mathbb{E}[\text{Tr}[A(k)_{j'j} \rho_A(j) A(k)_{jj'} \rho_A(j')]].
\end{equation}
What remains to be computed in the expression above  
Finally, $\mathbb{E}[\text{Tr}[A(k)_{j'j} \rho_A(j) A(k)_{jj'} \rho_A(j')]]$ is the expectation value of a functional of the fully asymmetric Haar random states $\{\rho_A(j)\}_j$. To obtain it, we can use the known results about the Haar average of a tensor product of unitary matrices, which can be obtained via Weingarten calculus. In particular, we need to use the two following identities \cite{w-78,cs-06}:
\begin{equation}
    \mathbb{E}[\rho_A(q)_{ab}] = \frac{1}{d_{A,q}} \delta_{ab},
\end{equation}
and
\begin{equation}\label{eq:weingarten_2}
    \mathbb{E}[\rho_A(q)_{ab}\rho_A(q)_{cd}] = \frac{1}{d_{A,q}(d_{A,q}d_{B,q}+1)}\big(\delta_{ad}\delta_{bc} + d_{B,q}\, \delta_{ab} \delta_{cd}\big).
\end{equation}
Moreover, Haar random states in different charge sectors are independent, i.e.~$\mathbb{E}[\rho_A(q)_{ab}\rho_A(q')_{cd}] = \mathbb{E}[\rho_A(q)_{ab}]\,\mathbb{E}[\rho_A(q')_{cd}]$.
Applying the previous results and after some straightforward algebra, we find
\begin{equation}
\label{expvtrace}
    \mathbb{E}[\text{Tr}[A(k)_{j'j} \rho_A(j) A(k)_{jj'} \rho_A(j')]] = \begin{cases}
        \frac{1}{d_{A,j}d_{A,j'}} \text{Tr}[A(k)_{jj'}A(k)_{j'j}], & j\neq j' \\
        \frac{(d_{B,j} \text{Tr}[A(k)_{jj}]^2 + d_{B,j}^2 \text{Tr}[A(k)_{jj}A(k)_{jj}])}{d_{A,j}d_{B,j}(d_{A,j}d_{B,j}+1)}, & j=j'
    \end{cases}.
\end{equation}
To keep the discussion clearer, we now only give the explicit expression of the traces $\text{Tr}[A(k)_{jj}]$ and $\text{Tr}[A(k)_{jj'}A(k)_{j'j}]$ appearing in Eq.~\eqref{expvtrace} and postpone their derivation to Sec.~\ref{subSM1}. They read
\begin{equation}
\label{trA}
    2^{\ell_A} \text{Tr}[A(k)_{jj}] = d_{A,j} d_{A,k},
\end{equation}
and
\begin{multline}
\label{TrAA}
    2^{2 \ell_A} \text{Tr}[A(k)_{jj'}A(k)_{j'j}] = d_{A,k} \sum_{m =0 }^{k-1} \binom{k}{m} \binom{\ell_A-k}{k-m} \\ \times \sum_{n = 0}^{\text{min}(2(k-m),j)}\,\, \sum_{n' = 0}^{\text{min}(2(k-m),j')} (-1)^{n+n'} \binom{2(k-m)}{n} \binom{2(k-m)}{n'}\binom{\ell_A-2(k-m)}{j-n} \binom{\ell_A-2(k-m)}{j'-n'}.
\end{multline}
To obtain the final result for the R\'enyi-2 entanglement asymmetry, $\mathbb{E}[\Delta S^{(2)}_A|_{Q_x}] = \mathbb{E}[\Delta S^{(2)}_A|_{Q_y}] \simeq -\log(\mathbb{E}[\text{Tr}{\rho_{A,Q_x}^2}]/\mathbb{E}[\text{Tr}{\rho_{A}^2}])$, we also need the average purity $\mathbb{E}[\text{Tr}{\rho_{A}^2}]$. Using the same techniques applied before, we have
\begin{equation}
\mathbb{E}[\text{Tr}{\rho_{A}^2}] = \sum_{q=0}^{\ell_A} \mathbb{E}[p_q^2] \, \mathbb{E}[\text{Tr}[\rho_A(q)^2]]
\end{equation}
and applying Eqs.~\eqref{eq:av_p_q_2} and~\eqref{eq:weingarten_2},
\begin{equation}
\label{avgpurity}
    \mathbb{E}[\text{Tr} \rho_A^2] = %\sum_{q=0}^{\ell_A} \mathbb{E}[p_q^2] \frac{d_{A,q} + d_{B,q}}{d_{A,q} d_{B,q} + 1} = 
    \frac{1}{d_M(d_M+1)}\sum_{q=0}^{\ell_A} d_{A,q} d_{B,q}(d_{A,q} + d_{B,q}).
\end{equation}
Putting all the pieces together, we obtain
\begin{equation}
    R\equiv\frac{\mathbb{E}[\text{Tr}[\rho_{A,Q_x}^2]]}{\mathbb{E}[\text{Tr} \rho_A^2]} = \frac{2^{-2 \ell_A}\binom{2 \ell_A}{\ell_A} \sum_{j} d_{B,j} d_{A,j}^2 + \sum_k \sum_{j,j'} d_{B,j} d_{B,j'} \text{Tr}[A(k)_{jj'}A(k)_{j'j}]}{ \sum_{j} d_{B,j} d_{A,j}^2 + \sum_{j} d_{B,j}^2 d_{A,j}}.
\end{equation}
Plugging Eq.~\eqref{TrAA} in the latter equation we can further simplify the triple sum 
\begin{equation}
\chi(L, \ell_A, M)= \sum_k \sum_{j,j'} d_{B,j} d_{B,j'} \text{Tr}[A(k)_{jj'}A(k)_{j'j}]
\end{equation}
appearing in the numerator as follows. Inserting Eq.~\eqref{TrAA} and  performing explicitly the sums in $j$ and $j'$,
\begin{equation}
     \chi(L, \ell_A, M) =2^{-2 \ell_A} \sum_k d_{A,k} \sum_{m=0}^k \binom{k}{m} \binom{\ell_A-k}{k-m} \Bigg[ \sum_{n= 0}^{2(k-m)}(-1)^n \binom{2(k-m)}{n} \binom{L-2(k-m)}{M-n}\Bigg]^2. 
\end{equation}     
The sum over $n$ can be rewritten as an ordinary hypergeometric function ${}_2 F_1(a,b,c;z)$,    
\begin{multline}
   \chi(L, \ell_A, M)=\\  2^{-2 \ell_A} \sum_k d_{A,k} \sum_{m=0}^k \binom{k}{m} \binom{\ell_A-k}{k-m} \binom{L-2(k-m)}{M}^2 {}_2 F_1(-2(k-m),-M,1-2(k-m)+L-M;-1)^2.
\end{multline}
Performing the change of variable  $m' = k-m$,
\begin{multline}
    \chi(L, \ell_A, M)=2^{-2 \ell_A} \sum_{k = 0}^{\ell_A} d_{A,k} \sum_{m'=0}^k \binom{k}{m'} \binom{\ell_A-k}{m'} \binom{L-2m'}{M}^2 {}_2 F_1(-2m',-M,1-2m'+L-M;-1)^2.
\end{multline}
Exchanging the order of the sums in $k$ and $m'$,
\begin{multline}
    \chi(L, \ell_A, M)=
    2^{-2 \ell_A} \sum_{m'=0}^
    {\ell_A}\binom{L-2m'}{M}^2 {}_2 F_1(-2m',-M,1-2m'+L-M;-1)^2\sum_{k=m'}^{\ell_A}\binom{\ell_A}{k} \binom{k}{m'} \binom{\ell_A-k}{m'}.
\end{multline}
 Finally, using the identity 
 \begin{equation}
 \sum_{k=m}^{\ell_A} \binom{\ell_A}{k}\binom{k}{m} \binom{\ell_A-k}{m} = 2^{\ell_A-2m} \binom{\ell_A}{m}\binom{\ell_A-m}{m} = 2^{\ell_A-2m} \frac{(2m)!}{m!^2}\binom{\ell_A}{2m},
 \end{equation}
 we get
\begin{equation}
    \chi(L, \ell_A, M)= 2^{-\ell_A} \sum_{m=0}^{\ell_A} \frac{2^{-2m} (2m)!}{(m!)^2} \binom{L-2m}{M}^2 \binom{\ell_A}{2m} {}_2 F_1(-2m,-M,1-2m+L-M;-1)^2,
\end{equation}
which is precisely the result stated in Eq.~(6) of the main text.

\subsection{Determination of Eqs.~\eqref{trA}-\eqref{TrAA} }\label{subSM1}

Let us now obtain Eqs.~\eqref{trA}~and~\eqref{TrAA}. As a first step, we rewrite the matrix elements $C_{qq'}^{mm'}$ as follows
\begin{equation}
    C_{qq'}^{mm'} = \bra{qm}e^{- i\frac{\pi}{4} \hat{Q}_{y,A}} \ket{q'm'} = \frac{1}{2^{\ell_A/2}} \bra{qm} \bigotimes_{j=1}^{\ell_A}\begin{pmatrix}
        1 & -1 \\
        1 & 1 
    \end{pmatrix} \ket{q'm'}.
\end{equation}
The elements of the basis $\{\ket{qm}\}$ can be explicitly written as the product states $\ket{qm}=\ket{b_1 \cdots b_{\ell_A}}$ where $b_j = 0,1$ and $\sum_j b_j = q$. Note that $m$ just labels all the possible arrangements of $q$ spins in the state $\ket{1}$ in a subsystem of $\ell_A$ sites.
Using this basis we have
\begin{equation}
\begin{split}
    2^{\ell_A/2} C_{qq'}^{mm'} &= \bra{b_1 \cdots b_{\ell_A}}\bigotimes_{j=1}^{\ell_A}\begin{pmatrix}
        1 & 1 \\
        1 & -1 
    \end{pmatrix} \ket{b_1' \cdots b_{\ell_A}'} = \\ &=  \prod_{i = 1}^{\ell_A} \big(\delta_{b_i,0} - (-1)^{b'_i} \delta_{b_i,1}\big) = \\
    &= \prod_{i = 1}^{\ell_A} (-1)^{\delta_{b_i,1} \Bar{b}'_i},
\end{split}
\end{equation}
where we have defined $\Bar{0} \equiv 1$ and $\Bar{1} \equiv 0$. In this representation, obtaining the value of $\text{Tr}[A(k)_{jj}]$ is immediate; in fact,
\begin{equation}
    \text{Tr}[A(k)_{jj}] = \sum_{m}\sum_{m'} C^{mm'}_{jj} C^{mm'*}_{jj} = \frac{1}{2^{\ell_A}} \sum_{m,m'} (-1)^{2 \delta_{b_i,1} \Bar{b}_i'} = \frac{1}{2^{\ell_A}} \sum_{m,m'} = \frac{1}{2^{\ell_A}} d_{A,j} d_{A,k}.
\end{equation}
In the same way, we can express $\text{Tr}[A(k)_{jj'}A(k)_{j'j}]$ as 
\begin{equation}
    2^{2 \ell_A} \text{Tr}[A(k)_{jj'}A(k)_{j'j}] = \sum_{\mathcal{C}} \prod_{i = 1}^{\ell_A} (-1)^{\delta_{b_i,1} \Bar{b}_i' + \delta_{b_i,1} \Bar{\Tilde{b}}_i'+
    \delta_{\Tilde{b}_i,1} \Bar{\Tilde{b}}_i' +
    \delta_{\Tilde{b}_i,1} \Bar{b}_i'
    },
\end{equation}
where by $\sum_{\mathcal{C}}$ we mean the sum over all possible configurations of $\{b_i,b_i',\Tilde{b}_i,\Tilde{b}_i'\}_{i = 1, \cdots,\ell_A}$ with the constraints $\sum_{i}b_i = \sum_{i} \Tilde{b}_i = k$, $\sum_i b_i' = j$, $\sum_{i} \Tilde{b}_i' = j'$. The terms appearing in the exponent can be factorised as follows
\begin{equation}
\label{trAApart}
 2^{2 \ell_A} \text{Tr}[A(k)_{jj'}A(k)_{j'j}] = \sum_{\mathcal{C}} \prod_{i=1}^{\ell_A} (-1)^{(\delta_{b_i,1} + \delta_{\Tilde{b}_i,1})(\Bar{b_i'}+\Bar{\Tilde{b}}_i')}.
 \end{equation}
We can use the following graphical notation to characterize each configuration. For the $i$-th qubit, we have four bits $\{b_i,b_i',\Tilde{b}_i,\Tilde{b}_i'\}$. We assign to these four bits the following diagram
\begin{equation}\label{eq:diagram}
\begin{tikzpicture}[scale=0.5,baseline=(current  bounding  box.center)]
\node at (-2,3.5) {$b_i$};
\node at (-2,2.5) {$\Tilde{b}_i$};
\node at (-2,1.5) {$b'_i$};
\node at (-2,0.5) {$\Tilde{b}'_i$};
\draw[->] (-1.5,2)-- (-0.25,2);
\foreach \x in {0, 1, 2, 3} {
    \draw[] (0,\x) rectangle (1,\x+1); 
}
% Ball in the first square
\fill[black,very nearly opaque] (0.5, 0.5) circle(0.25);
% Ball in the second square
%\fill[black] (1.5, 0.5) circle(0.3);
% Ball in the third square
%\fill[black] (2.5, 0.5) circle(0.3);
% Ball in the fourth square
%\fill[black] (3.5, 0.5) circle(0.3);
\end{tikzpicture}
\end{equation}
where
\begin{tikzpicture}[baseline = +0.5ex,scale=0.5]
    \draw[] (0,0) rectangle (1,1);
    \fill[black,very nearly opaque] (0.5, 0.5) circle(0.25);
\end{tikzpicture} $=1$ and \begin{tikzpicture}[baseline = +0.5ex,scale=0.5]
    \draw[] (0,0) rectangle (1,1);
\end{tikzpicture} $=0$. Each configuration $\mathcal{C}$ can then be thought as a $4\times\ell_A$ array of boxes filled with balls with the constraint that the sum of balls along the first and the second row is $k$, along the third row $j$ and along the fourth $j'$.

The crucial point in the computation of Eq.~\eqref{trAApart} is the parity of the value $(\delta_{b_i,1} + \delta_{\Tilde{b}_i,1})(\Bar{b_i'}+\Bar{\Tilde{b}}_i')$ for each diagram~\eqref{eq:diagram}. If the parity is odd, then the diagram contributes with a $-1$ to the product in \eqref{trAApart}, otherwise its contribution is a factor $1$. Hence we need to determine only the configurations with an odd contribution. These are given by the following set of four diagrams, which we denote as $F$,
\begin{equation}
\label{setF}
    \begin{tikzpicture}[baseline=(current  bounding  box.center)]
        \node at (2.25,-0.2) {$F=$};
        \draw[decoration={brace,mirror,raise=5pt,amplitude = 5pt},decorate, thick]
  (3.25,1.) -- node[right=6pt] {} (3.25,-1.5);
  \begin{scope}[scale = 0.5, shift = {(6.5,-2.5)}]
  \foreach \x in {0, 1, 2, 3} {
    \draw[] (0,\x) rectangle (1,\x+1); 
}
\fill[black,very nearly opaque] (0.5, 3.5) circle(0.25); 
\fill[black,very nearly opaque] (0.5, 1.5) circle(0.25);
  \end{scope}
  \node at (4,-0.25) {,};
  \begin{scope}[scale = 0.5, shift = {(8.5,-2.5)}]
  \foreach \x in {0, 1, 2, 3} {
    \draw[] (0,\x) rectangle (1,\x+1); 
}
\fill[black,very nearly opaque] (0.5, 2.5) circle(0.25);
\fill[black,very nearly opaque] (0.5, 0.5) circle(0.25);
  \end{scope}
   \node at (5,-0.25) {,};
  \begin{scope}[scale = 0.5, shift = {(10.5,-2.5)}]
  \foreach \x in {0, 1, 2, 3} {
    \draw[] (0,\x) rectangle (1,\x+1); 
}
\fill[black,very nearly opaque] (0.5, 3.5) circle(0.25);
\fill[black,very nearly opaque] (0.5, 0.5) circle(0.25);
  \end{scope}
   \node at (6,-0.25) {,};
  \begin{scope}[scale = 0.5, shift = {(12.5,-2.5)}]
  \foreach \x in {0, 1, 2, 3} {
    \draw[] (0,\x) rectangle (1,\x+1); 
}
\fill[black,very nearly opaque] (0.5, 1.5) circle(0.25);
\fill[black,very nearly opaque] (0.5, 2.5) circle(0.25);
  \end{scope}
 \draw[decoration={brace,raise=5pt,amplitude = 5pt},decorate, thick](6.75,1.) -- node[right=6pt] {} (6.75,-1.5);
 \node at (7.25,-0.25) {.};
 \end{tikzpicture}
\end{equation}
 The contribution to the sum in~\eqref{trAApart} of each configuration is one if the number of columns belonging to $F$ is even or minus one if the number of columns belonging to $F$ is odd. Let $N_e$ ($N_o$) be the number of configurations with an even (odd) number of $F-$columns, Eq.~\eqref{trAApart} can then be rewritten as
\begin{equation}
\label{trAAinterm}
\begin{split}
    2^{\ell_A} \text{Tr}[A(k)_{jj'}A(k)_{j'j}] &= N_e-N_o\\
    &=d_{A,k}^2d_{A,j}d_{A,j'} - 2 N_o.
\end{split}    
\end{equation}
In the second equality, we took into account that the total number of configurations is $N_e+N_o=d_{A,k}^2d_{A,j}d_{A,j'} = \binom{\ell_A}{k}^2\binom{\ell_A}{j}\binom{\ell_A}{j'}$. We can then eliminate $N_e$ in Eq.~\eqref{trAAinterm} and only $N_o$ remains unknown. Its computation boils down to solving the combinatorial problem of determining in how many ways we can arrange $2k+j+j'$ balls in a $4\times\ell_A$ array of boxes with the constraints that the first two rows contain $k$ balls each, the third $j$ and the fourth $j'$ and such that the number of columns belonging to the set $F$ in Eq.~\eqref{setF} is odd.

The value $N_o$ is, by definition, invariant under any permutation of the columns of the $4\times \ell_A$ array, hence, we can fix the arrangement of the $k$ balls in the first row and multiply by $d_{A,k}$ the remaining counting. We then sum over the possible arrangements of the $k$ balls in the second row by ordering the sum based on the number of $m$ balls that share the same column. We make this choice as it will simplify the problem of counting the remaining configurations. At this point, we have
\begin{equation}
\label{No_step}
 N_o = d_{A,k} \sum_{m = 0}^k \binom{k}{m} \binom{\ell_A-k}{k-m} \cdot \mathcal{N}(k,m,j,j'),
\end{equation}
where $\mathcal{N}(k,m,j,j')$ is the number of configurations with odd $F-$columns but with a fixed arrangement for the first two rows such that $m$ columns contain two balls in the first two boxes. To determine $\mathcal{N}(k,n,j,j')$, we notice that an $F-$column can be created only among those columns in which one of the two first rows is occupied but not both. There are $2(k-m)$ of these columns. Observe now that, if we distribute $n$ and $n'$ balls in the third and fourth rows only among the $2(k-m)$ columns, then the number of $F-$columns is given by $n+n'$ (the possibility of having an overlap between third and fourth column does not change the parity). For this reason, $\mathcal{N}(k,n,j,j')$ is given by
\begin{equation}
\begin{split}
    &\mathcal{N}(k,n,j,j') = \\&=\sum_{n=0}^{\text{min}(2(k-m),j)} \sum_{n'=0}^{\text{min}(2(k-m),j')} \delta_{n+n',2p+1} \binom{2(k-m)}{n}\binom{2(k-m)}{n'}\binom{\ell_A-2(k-m)}{j-n}\binom{\ell_A-2(k-m)}{j'-n'},
\end{split}
\end{equation}
with $p\in \mathbb{Z}$. Inserting this result into \eqref{No_step}, we obtain the expression for $N_o$. If we write the Kronecker delta as $(1-(-1)^{n+n'})/2$, then it is easy to see that the sum without alternating sign and the term $d_{A,k}^2d_{A,j}d_{A,j'}$ mutually cancel in Eq.~\eqref{trAAinterm} and we finally arrive at the result in Eq.~\eqref{TrAA}.

\section{Additional numerical results}\label{sec:AddNumerics}
\begin{figure*}[t]
\centering
\includegraphics[width=0.48\textwidth]{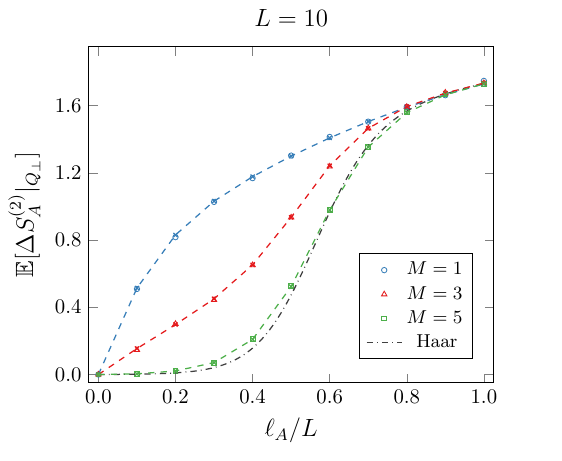}
\includegraphics[width=0.48\textwidth]{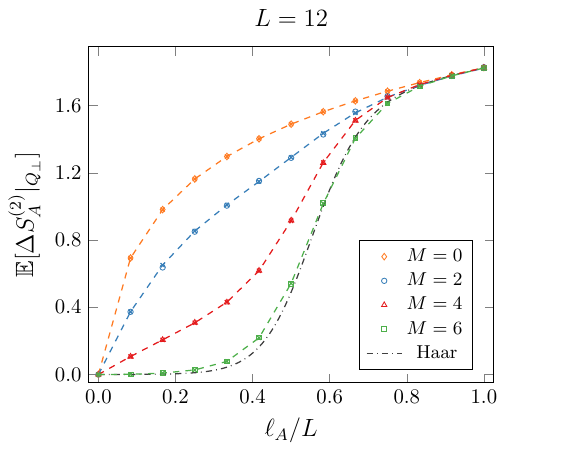}
  \caption{\label{fig:sampleU1Haar}
  Check of the analytical result~(7) for the average R\'enyi-2 entanglement asymmetry $\mathbb{E}[\Delta S^{(2)}_A|_{Q_\perp}]$ for a charge $Q_\perp$ orthogonal to the conserved one. The prediction of Eq.~(7) is is represented as crosses joint by a dashed line which is only meant to guide the eye. We plot $\mathbb{E}[\Delta S^{(2)}_A|_{Q_\perp}]$ as a function of the ratio $\ell_A/L$ for different charge sector $M$ of the full Hilbert space and total system sizes $L=10$ (left panel) and $12$ (right panel). The symbols are the exact average asymmetry over a set of $\sim 200$ states sampled from the ensemble of $U(1)$-symmetric Haar random states in the charge sector $M$. These data were calculated without using the self-averaging approximation, in contrast to the analytical result~(7). The number of random states considered is enough to make the statistical error on the mean smaller than the marker size.}
\end{figure*}

In this section, we first numerically check Eq.~(7) in the main text. To this end, we sample random states belonging to the ensemble of $U(1)$-symmetric Haar random states in the same charge sector $M$. We then numerically compute their entanglement asymmetry and average over the finite sample of results. What we obtain is an estimate of the expectation value $\mathbb{E}[\Delta S^{(2)}_A|_{Q_{\perp}}]$. In Fig.~\ref{fig:sampleU1Haar}, we compare the numerical results with Eq.~(7), finding an excellent agreement. Moreover, note that this further validates the self-averaging approximation $ \mathbb{E}[\log \text{Tr}[\rho_{A,Q}^2]]\simeq \log \mathbb{E}[ \text{Tr}[\rho_{A,Q}^2]]$ applied to derive Eq.~(7) from Eq.~(5).

\begin{figure}[t]
\includegraphics[width=0.49\textwidth]{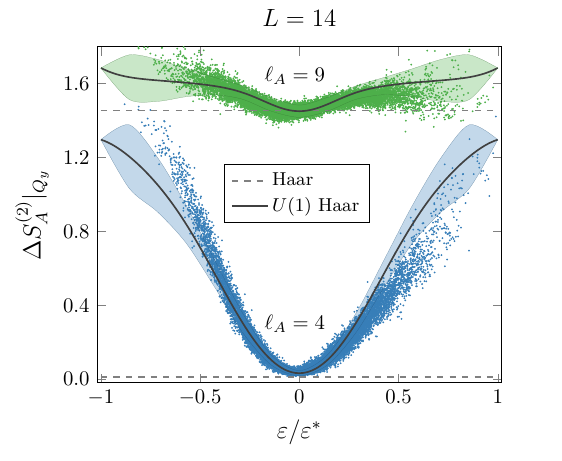}
\includegraphics[width=0.49\textwidth]{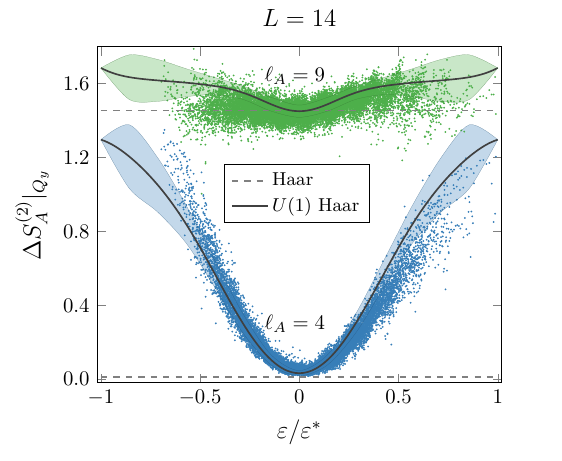}
\centering
\caption{The symbols are the R\'enyi$-2$ entanglement asymmetry for the charge $Q_y$ of each eigenstate of the Hamiltonian~\eqref{modelsm1} (left panel) and~\eqref{modelsm2} (right panel), with $L = 14$ sites and parameters $g = 1.1$, $h = 0.35$ and $\Delta=2$. We study two different subsystems with length $\ell_A = 4$ (blue points) and $\ell_A = 9$ (green points). The energy density of the eigenstates $\varepsilon$ has been rescaled with $\varepsilon^* = 1.35$ (left panel) and $\varepsilon^*=1.33$ (right panel), as explained in the main text. The black full line is the analytical prediction~(7) obtained for the ensemble of $U(1)$-symmetric random states with conserved charge orthogonal to $Q_y$.  The blue and green shaded regions correspond to the confidence interval $\mathbb{E}[\Delta S^{(2)}_A]\pm 3 \sigma$, where the variance $\sigma$ has been estimated numerically by sampling the ensemble of random states. The grey dashed line is the prediction~(8) of the standard Haar unitary ensemble, which should give an approximate prediction for only the mid-spectrum eigenstates.}
\label{fig:asymfullspectrumSM1} 
\end{figure}

As a complement to the results of the main text, we have studied the entanglement asymmetry in other chaotic local interacting spin-$1/2$ Hamiltonians and verified the universality of our predictions using the $U(1)$-symmetric random states and the expected form of the associated conserved charge. We consider the following two Hamiltonians,
\begin{equation}
\label{modelsm1}
    H_1 = \sum_{j=1}^{L-1}\left( \frac{1}{2}\sigma^z_j \sigma^z_{j+1}+\frac{1}{2}\sigma^z_j \sigma^z_{j+2}\right)+ \sum_j\left(g \sigma^x_j + h \sigma^z_j\right),
\end{equation}
\begin{equation}
\label{modelsm2}
    H_2 =\frac{1}{4} \sum_{j=1}^{L-1}\left(\sigma^x_j \sigma^x_{j+1}+\sigma^y_j \sigma^y_{j+1}+\Delta \sigma^z_j\sigma^z_{j+1}\right)+\sum_{j}\left(g \sigma^x_j + h \sigma^z_j\right).
\end{equation}
In the Hamiltonian in Eq.~\eqref{modelsm1}, we have added a next-to-nearest neighbor interaction compared to Eq.~(2). In Eq.~\eqref{modelsm2}, the interacting part is the integrable XXZ spin-$1/2$ chain, but the external magnetic fields along the $x$ and $z$ axis break integrability. In both cases, we introduce the boundary fields $h_1 \sigma^z_1,\,h_L \sigma^z_L$ with $h_1 = 1/4,\, h_L = -1/4$. Each model has thus no symmetry except for energy conservation and we expect that, for generic values of the couplings $g$, $h$ of order $O(1)$, they are chaotic. We choose again $g = 1.1$ and $h = 0.35$. Fig.~\ref{fig:asymfullspectrumSM1} is identical to Fig.~1 in the main text but plotting the asymmetry of the eigenstates of~\eqref{modelsm1} (left panel) 
and~\eqref{modelsm2} with respect to the charge $Q_y$. The solid curves are the average asymmetry~(7) of the random state ensemble with a $U(1)$ symmetry generated by the charge $gQ_x+hQ_z$. The total charge $M$ of the random states is related to the eigenstate energy density $\varepsilon$ through the identity in Eq.~(3). 
We remark that, in this case, the parameters $g$ and $h$ (and $\Delta$ for the model \ref{modelsm2}) do not necessarily correspond to the maximally chaotic point (at least in the sense of Ref.~\cite{rjk-24}). Nonetheless, we find a reasonable agreement between the numerics and the analytical formula~(7).
We observe a modulation in the numerics in the right panel of Fig.~\ref{fig:asymfullspectrumSM1}. A similar modulation occurs in the density of states. This effect likely indicates that, for the parameters chosen, the model is not far enough from an integrable point \cite{ab-14}.

\end{document}